\documentclass[review,12pt]{elsarticle}

\usepackage{booktabs}
\usepackage{graphicx}
\usepackage[utf8]{inputenc}
\usepackage{float}
\usepackage{xcolor}
\usepackage{soul}
\usepackage{xurl}
\usepackage{lineno}
\usepackage{amsmath}

\journal{Applied Ergonomics}

\begin{document}
\begin{frontmatter}

\title{Multisensor Measurement of Train Driver Mental Fatigue: From Simulation to Reality}

\author[inst1]{Esther Bosch\corref{cor1}}
\author[inst1]{Rebecca Kruschka}
\author[inst1]{David Schackmann}
\author[inst2]{Stephanie Hoyer}
\author[inst3]{Wolfgang Kilian}
\author[inst3]{Stefan Schwanitz}
\author[inst4]{Anneke Hamann}

\cortext[cor1]{corresponding author: esther.bosch@dlr.de}

\affiliation[inst1]{organization={German Aerospace Center, Institute of Transportation Systems},city={Braunschweig}, country={Germany}}
\affiliation[inst2]{organization={ISB Ingenieurgesellschaft für Sicherungstechnik und Bau mbh},city={Annaberg-Buchholz}, country={Germany}}
\affiliation[inst3]{organization={Chemnitz University of Technology, Department of Sports Equipment and Technology},city={Chemnitz}, country={Germany}}
\affiliation[inst4]{organization={German Aerospace Center, Institute of Flight Guidance},city={Braunschweig}, country={Germany}}

\begin{abstract}
Increasing automation in rail transport shifts the train driver's role from active control to prolonged supervisory monitoring. This creates conditions for mental fatigue (MF) and reduced vigilance. Despite the safety relevance of this issue, evidence on the feasibility and robustness of physiological indicators of MF under operational rail conditions remains limited. Most prior work relies on simulators or lab studies.

The present study investigated multiple subjective, physiological, and behavioral indicators of MF in professional train drivers across two complementary settings: a high-fidelity train simulator (\textit{n}=14) and a real-world rail environment (\textit{n}=6). To our knowledge, this is the first study to deploy a full multisensor battery under actual train operating conditions. In both settings, a standardized protocol was used comprising a baseline drive, a one-hour auditory n-back task as an MF induction procedure, and a second drive. 

Heart rate variability and breathing rate showed consistent and theoretically expected changes across both environments, suggesting reduced physiological arousal following the fatigue induction task. In contrast, EEG-based frontal theta power and parietal alpha and beta power, electrodermal activity, blink duration, and behavioral indicators did not show clear mental fatigue-related patterns. Real-world data collection revealed substantial technical challenges related to vibration, sensor connectivity, and concurrent high-frequency data acquisition. 

These findings suggest that autonomic indicators, particularly HRV and breathing rate, represent the most promising and ecologically robust measures for operational fatigue monitoring in train drivers. However, neurophysiological measures require further validation under realistic conditions before deployment in driver monitoring systems, and larger samples are needed to confirm these preliminary patterns.

\end{abstract}



\end{frontmatter}



\section{Introduction}

Higher levels of automation in railway operation can improve operational efficiency while also introducing human-factors challenges \cite{brandenburger2021task}. In particular, the shift in the train driver's role from manual control to system monitoring illustrates that, at intermediate levels of automation, reliable human performance remains essential, although the conditions under which it must be maintained change substantially. The Grades of Automation (GoA) framework describes levels of automation in railway operation, ranging from on-sight train operation (GoA0) to unattended train operation (GoA4) \cite{uitp2012metro}. In GoA1 systems, train drivers are predominantly engaged in manual control, whereas in GoA2 systems, acceleration and braking are automated while the driver remains responsible for door operation, system supervision, and intervention in the case of unexpected events \cite{uitp2012metro}. This transition towards prolonged monitoring with reduced active engagement reflects Bainbridge's Ironies of Automation. According to Bainbridge, increasing automation reduces human activity while still requiring reliable system monitoring. Such monitoring is difficult to sustain over extended periods due to limited cognitive resources \cite{BAINBRIDGE1983129}. This concern is particularly relevant in GoA2 operation, where the higher degree of automation may decrease train drivers' mental workload \cite{dlr119746} and, in combination with the monotonous nature of continuous monitoring, contribute to the onset of task-induced mental fatigue (MF) \cite{brandenburger2021task, FILTNESS201712, DOBSON20152913}. Accordingly, MF represents a relevant operator state that may affect train drivers' monitoring performance.

In the context of train drivers' monitoring tasks, MF is particularly relevant for railway safety. MF can be broadly understood as a task-induced, acute, non-pathological state that lies on a continuum between a restful, relaxed wakeful state and sleep \cite{10.3389/fnrgo.2025.1673268, Grandjean175}. It has been shown to impair cognitive and executive functioning, including sustained attention, as reflected in diminished accuracy and longer reaction times \cite{guo2016impairing}. MF has also been associated with reduced situation awareness \cite{zhou2023situationawareness}, deficits in action monitoring and error detection \cite{8616098, BOKSEM2006123}, and diminished inhibitory control \cite{BOKSEM2005107}. These functions are essential in safety-critical monitoring tasks, in which operators must detect irregularities and respond promptly to unexpected events. Evidence from other transportation domains further illustrates the safety relevance of MF. In car driving, fatigue may contribute to reduced alertness, as indicated by slower reaction times when braking and steering in emergency situations \cite{saxby2013active}. Similarly, in aviation, MF has been associated with reduced monitoring and signal detection performance \cite{5377cc41554b4171bdfd3a4cddd48a01, 8616098}, lower situation awareness, and impaired responses to unexpected events in safety-critical situations \cite{zhou2023situationawareness, 10.3389/fnrgo.2025.1673268}. In the railway context, MF may contribute to safety-critical performance decrements, including distracted driving and passing stop signals, such as signals passed at danger (SPADs) \cite{FILTNESS201712}. Despite its relevance for railway safety, the objective measurement of MF remains challenging.

MF is a subjective and multidimensional construct that cannot be observed directly and is therefore assessed through indirect indicators \cite{bjegojevic2021physiological}. It is commonly assessed using multimodal approaches that combine subjective self-reports, physiological and behavioral signals, and performance-related indicators \cite{kunasegaran2023understanding, schampheleer2025current}. In transportation research, multimodal approaches to MF assessment have been widely applied in road traffic, particularly in monotonous and partially automated driving situations. These studies have used a range of sensor approaches, including subjective scales \cite{saxby2013active}, self-reports and eye-tracking measures, driving-dynamics characteristics \cite{app13021200}, (neuro)physiological measurements \cite{brainsci15091001}, and camera-based head, gaze, and eyelid parameters \cite{9031734}. MF has also been investigated in aviation, for example using neurophysiological measurements during simulated flights \cite{hamann2023assessing}, heart rate variability \cite{guo2025assessment}, Electroencephalography (EEG) and heart rate variability \cite{chen2026assure}, and heart rate variability and eye tracking \cite{qin2021detection}. In the railway context, previous studies have primarily examined fatigue- or vigilance-related states using physiological and behavioral indicators, including EEG \cite{zhang2017design}, eye-tracking and heart-rate measures \cite{407d5eb2652e4f9280fbf09bd4a91982}, ECG \cite{ma2020study}, electrodermal activity \cite{dorrian2008driver}, and self-reported sleepiness \cite{bs13100788}. Respiration rate has also been discussed as a relevant indicator for detecting MF in broader fatigue research \cite{10.3389/fphys.2021.790292}, but appears less established in assessing MF in train drivers. Despite numerous empirical studies on detecting MF in drivers and the promising application of various sensor approaches \cite{9896910, LU2022106830}, there is still no generally accepted standard for quantifying MF or for selecting and combining suitable sensor systems \cite{9031734, wang2024review}. In one of the few studies conducted directly on real train operations, Song et al. measured EEG, ECG, Electrodermal Activity (EDA), photoplethysmography, and respiration in locomotive engineers during actual train driving, finding that skin conductance response increased in sections requiring greater body movement, alongside increases in heart rate and EEG beta activity around train stops and tunnel sections that reflected heightened mental arousal and tension \cite{song2014physiological}. Overall, compared with the automotive and aviation domains, the evidence base for multimodal and sensor-based MF assessment in the railway context remains limited \cite{https://doi.org/10.1155/2022/8328077}, with no studies applying the full range of available sensor modalities within a single rail study.

MF assessment in operational environments remains challenging, as strict requirements regarding non-intrusiveness, robustness, and user acceptance constrain deployment. Eye-tracking and camera-based systems are highly dependent on lighting conditions, and in the train driver context, solar glare, dashboard reflections \cite{10897716, 10.3389/ffutr.2025.1677442}, and rapid light transitions, such as when entering or exiting tunnels \cite{hassan2026intelligent}, can impair vision-based measurement accuracy. Train-cab vibration and seat dynamics may further induce residual upper-body or head motion \cite{smith2006vibration, STEIN2008384, nocentini2025graphbasedonlinemonitoringtrain}, interfering with both vision-based and physiological recordings \cite{9057686, s24196363, 10.3389/fnins.2024.1328704}. Cab layout also constrains sensor placement, and sensors must remain non-intrusive and comfortable without obstructing vision or restricting movement. For instance, EEG is effective for measuring MF \cite{KAR2010297} but multi-channel setups may be perceived as uncomfortable or burdensome by drivers \cite{s23198171}, while continuous camera- or eye-based monitoring can raise data privacy concerns \cite{kneffel2025nutzerakzeptanz}. These challenges highlight the need to evaluate both the sensitivity of individual MF-related parameters and the feasibility of multisensor approaches under ecologically valid railway conditions.

Against this background, sensor-based MF assessment in train drivers remains limited: the full range of operationally viable sensors has rarely been evaluated together, and the sensitivity, feasibility, and robustness of individual indicators under ecologically valid railway conditions remain insufficiently understood. The present study addresses this gap by investigating multiple subjective, physiological, and behavioral indicators of MF in train drivers across two complementary settings: a high-fidelity train simulator and real-world rail operation. By comparing results across controlled and operational conditions, the study aims to identify which behavioral and physiological indicators are sensitive to MF-related changes, and determine which remain sufficiently robust, feasible, and interpretable under the constraints of real-world rail environments. The study thereby contributes to the evaluation of multisensor approaches for MF assessment in the railway context and supports the further development of sensor-based MF detection in train drivers.

\section{Background}

\subsection{Mental Fatigue}

A clear conceptualisation of MF is essential for understanding its potential impact on train drivers' performance and railway safety. However, there is currently no universally accepted definition of MF or fatigue more generally \cite{PHILLIPS201548}, which complicates comparisons between empirical studies. The present paper therefore adopts Grandjean's \cite{Grandjean175} traditional and widely cited definition of MF. He describes MF as a sensation of weariness characterised by impaired activity, feelings of inhibition, and a reduced willingness to engage in mental or physical effort. He further situates MF among a range of intermediate functional states between alarm and sleep. Boksem and Tops explain this through cost-benefit evaluation of task effort \cite{BOKSEM2008125}. MF develops gradually during sustained task engagement \cite{CHARBONNIER201691, BOKSEM2008125}. Sleepiness differs from MF in its underlying mechanisms \cite{HU2020173}. Sleepiness stems from circadian disruption and prior sleep loss \cite{balkin2011differentiation}. Drowsiness can vary within seconds \cite{johns2008new, BORGHINI201458}, while MF accumulates more slowly than drowsiness \cite{CHARBONNIER201691}.

In the context of train driving, passive MF is particularly relevant. Whereas active MF arises from continuous perceptual-motor adjustments to task demands, passive MF develops during extended monitoring periods characterised by monotony and cognitive underload \cite{desmond2000active}. Evidence from automated road driving indicates that such low-demand conditions promote passive MF, accompanied by reduced task engagement, increased mind wandering, slower responses to unexpected hazards, and higher collision risk \cite{saxby2013active, KORBER20152403}. Although obtained in simulated road driving, these findings may extend to GoA2 train operation, where drivers similarly assume a supervisory role while retaining responsibility for detecting and responding to unexpected events. Consistent with this, Brandenburger et al. found that GoA2 was associated with lower workload but higher reported task-induced fatigue than GoA3, a pattern linked to the monotony and reduced cognitive demand of GoA2 operation \cite{brandenburger2021task}. Given the low-demand monitoring and supervisory conditions associated with GoA2 operation, the present study focuses particularly on the detection of passive MF.

\subsection{Traditional Mental Fatigue Monitoring and Its Limitations}
MF has been commonly examined through related performance constructs, particularly vigilance and sustained attention, as MF has been associated with reduced performance in tasks requiring these functions \cite{Smith17112019, BOKSEM2005107, bjegojevic2021physiological}. Vigilance refers to the ability to maintain attention and successfully detect relevant signals over prolonged periods \cite{brainsci9080178}. In this context, vigilance decrement, defined as a reduced likelihood of identifying rare but relevant events as time-on-task progresses during continuous stimulus monitoring \cite{mackworth1948breakdown}, represents a well-studied manifestation of fatigue-related performance decline and has been described as closely related to MF \cite{OKEN20061885, CHARBONNIER201691}.

Historically, vigilance in train driving has been addressed through technical vigilance monitoring devices, such as dead-man pedals or periodic acknowledgment systems, which require drivers to perform simple motor responses to confirm wakefulness. Early experimental and operational investigations demonstrated that these systems are limited in their ability to detect genuine reductions in vigilance, as drivers can develop highly automated response patterns that allow correct system operation even when overall vigilance and situation awareness are degraded \cite{whitlock2017driver, foot2008questions, berdal2024towards}. This limitation has motivated research into more direct and objective measures of operator state that are less susceptible to behavioral adaptation and training effects. For example, multimodal non-invasive measures could provide a promising indicator of MF \cite{HU2020173}.

\subsubsection{Eye Tracking Measures}

Eye tracking offers a set of indicators for fatigue, centered on eyelid closure, blink dynamics, and gaze behavior. Task-related fatigue elicits distinguishable saccade-, fixation-, and blink-based responses \cite{hu2021exploration}. Bodala et al. combined EEG and eye-tracking and support the validity of ocular measures, showing that vigilance changes are reflected jointly in EEG indices (e.g., frontal theta, parietal alpha) and oculomotor measures such as saccade velocity, amplitude, and blink rate, with significant correlations between the two modalities \cite{bodala2016eeg}.
Among these measures, eye-blink parameters stand out as a particularly feasible option for operational monitoring. In an on-road instrumented-vehicle study, blink-based metrics, including prolonged eyelid closure duration, total blink duration, and the amplitude-to-velocity ratio of eyelid movements, reliably detected drowsiness-related driving impairment under severe sleep deprivation, matching real driving outcomes such as out-of-lane events and early drive terminations \cite{shekari2019eye}. Because blink dynamics can be captured with comparatively simple, non-intrusive camera-based eye tracking, they can be a practical and scalable candidate for vigilance monitoring in operational rail settings.

\subsubsection{Physiological Measures}

Electroencephalography (EEG) has been one of the most frequently studied indicators of MF, with numerous rail-focused studies reporting promising results using a limited number of channels \cite{fan2022types, fan2021detection, zhang2017design, jap2011comparing}. Several simulator-based studies have demonstrated high classification accuracies for car driver or aircraft pilot fatigue detection using wearable EEG systems, often combined with machine learning approaches (e.g., \cite{BORGHINI201458}). Reported indicators for mental fatigue commonly include increases in frontal theta power and changes in parietal alpha-band and beta-band activity \cite{hamann2023assessing}. However, these findings require careful interpretation. Many studies rely on small samples, controlled simulation environments, and relatively short recording durations. Neuroergonomics research in aviation and automotive domains has shown that EEG markers commonly associated with workload or fatigue are highly sensitive to task characteristics, contextual factors, and confounding influences, and that their validity can degrade substantially outside tightly controlled laboratory conditions \cite{dehais2020neuroergonomics}.

Autonomic measures such as heart rate variability (HRV), respiration, and electrodermal activity (EDA) have been widely used as indicators of MF, stress, and vigilance across transportation and industrial domains. Electrocardiogram (ECG)-based measures in particular emerged as among the most frequently applied physiological signals for monitoring mental workload and fatigue across laboratory and field settings \cite{lal2001critical, charles2019measuring}. Reviews of operator-state monitoring consistently note that while no single physiological measure discriminates mental workload or fatigue reliably across all task types, cardiac measures benefit from comparatively simple, non-intrusive acquisition that makes them attractive for continuous, wearable-based monitoring \cite{charles2019measuring}. Recent HRV-based approaches have moved toward real-time applicability. A sliding-window HRV-based machine-learning method demonstrated that vigilance fluctuations during a sustained psychomotor task could be tracked in near real time by associating HRV features with concurrent behavioral performance metrics, addressing earlier limitations related to lengthy feature-extraction windows and reliance on subjective benchmarks \cite{xie2025tracking}. 

Respiration-based measures have likewise been explored as a non-invasive complement to cardiac and EEG-based approaches. For example, drowsiness detection systems incorporating respiration rate extracted via non-contact sensing (alongside other physiological channels) have demonstrated feasibility for practical, low-burden driver monitoring \cite{siddiqui2021non}. 

EDA reflects sympathetic nervous system activation via sweat gland activity and has been explored as a further non-invasive autonomic indicator of arousal, stress, and fatigue \cite{lal2001critical, charles2019measuring}. However, evidence on the specific sensitivity of EDA to MF, as opposed to general arousal or workload, remains mixed: a recent comparison of neurophysiological correlates across monotonous and cognitively demanding simulated driving conditions found that while an EEG-based drowsiness index remained sensitive to fatigue under both conditions, and HRV decreased mainly with higher cognitive demand, EDA showed no clear sensitivity to fatigue-related changes \cite{dello2025analysis}. This suggests that EDA's diagnostic value for MF may be limited relative to cardiac and neural measures.

Taken together, autonomic and respiratory measures may offer a pragmatic compromise between sensitivity to fatigue-related state changes and the robustness and ease of deployment required for operational, real-world monitoring, though, as with EEG and eye-tracking approaches, further validation under authentic rail operating conditions remains necessary.

\subsubsection{Behavioral Measures}

Behavioral indicators such as reaction times, vigilance task performance, and control inputs have likewise been explored as markers of fatigue and reduced attention \cite{charles2019measuring, lal2001critical}. Vigilance research has traditionally relied on such performance-based measures, since sustained-attention tasks were historically viewed as low-demand assignments for which simple detection or response accuracy would suffice as an indicator of operator state \cite{warm2008vigilance}. However, previous work has also shown that vigilance is highly task-type specific, with the sensitivity and workload characteristics of performance-based measures varying considerably depending on the structure and demands of the underlying task \cite{warm2008vigilance}. This finding was established across domains such as surveillance, cockpit monitoring, air traffic control, and the supervisory control of automated systems \cite{warm2008vigilance}. Moreover, drivers can continue to respond correctly on simple secondary or vigilance tasks through automatized motor patterns even while neurophysiological indicators reveal clear reductions in underlying vigilance, meaning that intact behavioral performance does not necessarily reflect a preserved underlying attentional state. Taken together, these limitations suggest that behavioral measures alone are unlikely to provide a complete or sufficiently early picture of fatigue and vigilance decrement. This motivates the complementary use of physiological measures capable of tracking the operator's internal state continuously and independently of discrete task responses.

Overall, the literature indicates that no single indicator provides a comprehensive or universally reliable measure of MF or vigilance in train drivers. Neurophysiological, autonomic, eye-based, and behavioral measures each offer complementary insights but differ substantially in their sensitivity, robustness, intrusiveness, and practical feasibility. Recent reviews and applied studies increasingly advocate for multisensor approaches combined with a critical evaluation of ecological constraints \cite{rivas2025combining, song2014physiological, dello2025analysis}. Rather than optimizing classification accuracy under idealized conditions, there is a need for empirical evidence on which indicators remain interpretable and usable in real-world rail environments. The present study builds on this work by systematically examining multiple fatigue-related indicators across simulator and real-world train driving, with a particular focus on feasibility, robustness, and practical suitability for applied rail ergonomics.

Therefore, this paper aims to answer the following Research Questions:

\begin{itemize}
    \item Mental Fatigue Induction: Do subjective indicators show n-back-induced MF changes in simulator and real-world conditions?
    \item Indicator Sensitivity: Which behavioral and physiological measures (heart rate variability, respiration, electrodermal activity, eye-blink metrics, EEG measures) show sensitivity to fatigue-related changes in train drivers?
    \item Ecological Robustness: Which indicators remain stable and interpretable under real-world rail operating conditions, considering factors such as movement, vibration, and operational constraints?
    \item Practical Feasibility: How do drivers perceive the comfort and usability of the employed sensors?
\end{itemize}

Based on the expected directions of change, we formulate the following hypotheses regarding physiological, behavioral, and subjective indicators of MF:

\begin{itemize}
\item Subjective fatigue manipulation check: longer time on task is expected to be associated with higher self-reported fatigue.
\item Electroencephalogram: MF is expected to be associated with increased frontal theta and parietal alpha and beta power \cite{hamann2023assessing}.
\item Electrocardiogram: MF is expected to be associated with a higher heart rate variability \cite{matuz2021enhanced}.
\item Electrodermal Activity: MF is expected to be associated with a reduction in the number of skin conductance response (SCR) peaks \cite{posada2017sleep}.
\item Blinking: MF is expected to be associated with longer blink duration \cite{bafna2021mental}.
\item Breathing: MF is expected to be associated with a slower breathing rate \cite{grassmann2016respiratory}.

\item Reaction time: MF is expected to be associated with longer reaction times.
\item Dead man's handle: MF is expected to be associated with a higher number of missed dead man's handle activations.

\end{itemize}

\section{Materials and Method}
\begin{figure}
\centering
    \includegraphics[width=\textwidth]{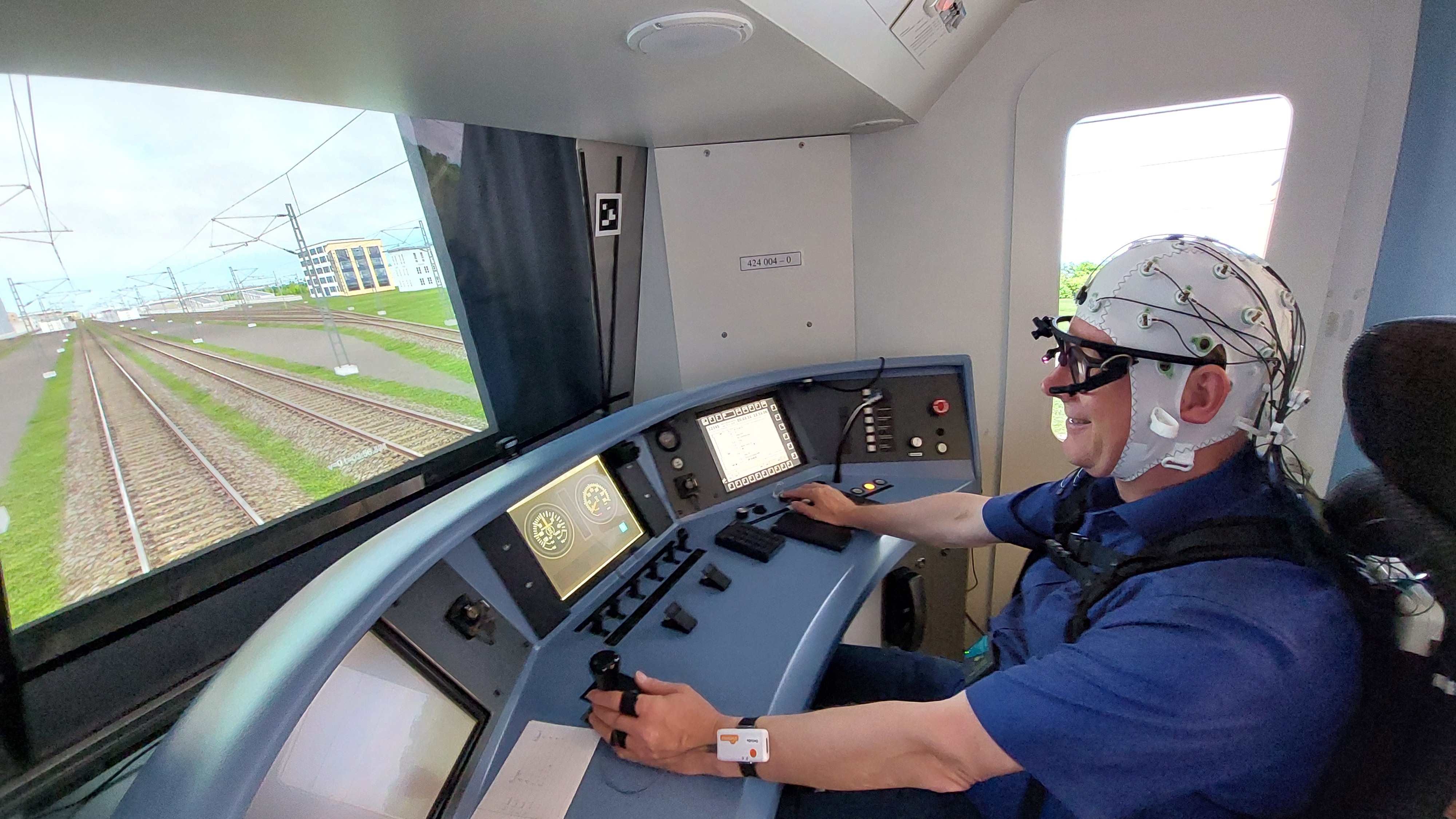}
    \caption{Train driving cab simulator.}
    \label{fig:first}
\end{figure}

\begin{figure}{}
    \includegraphics[width=\textwidth]{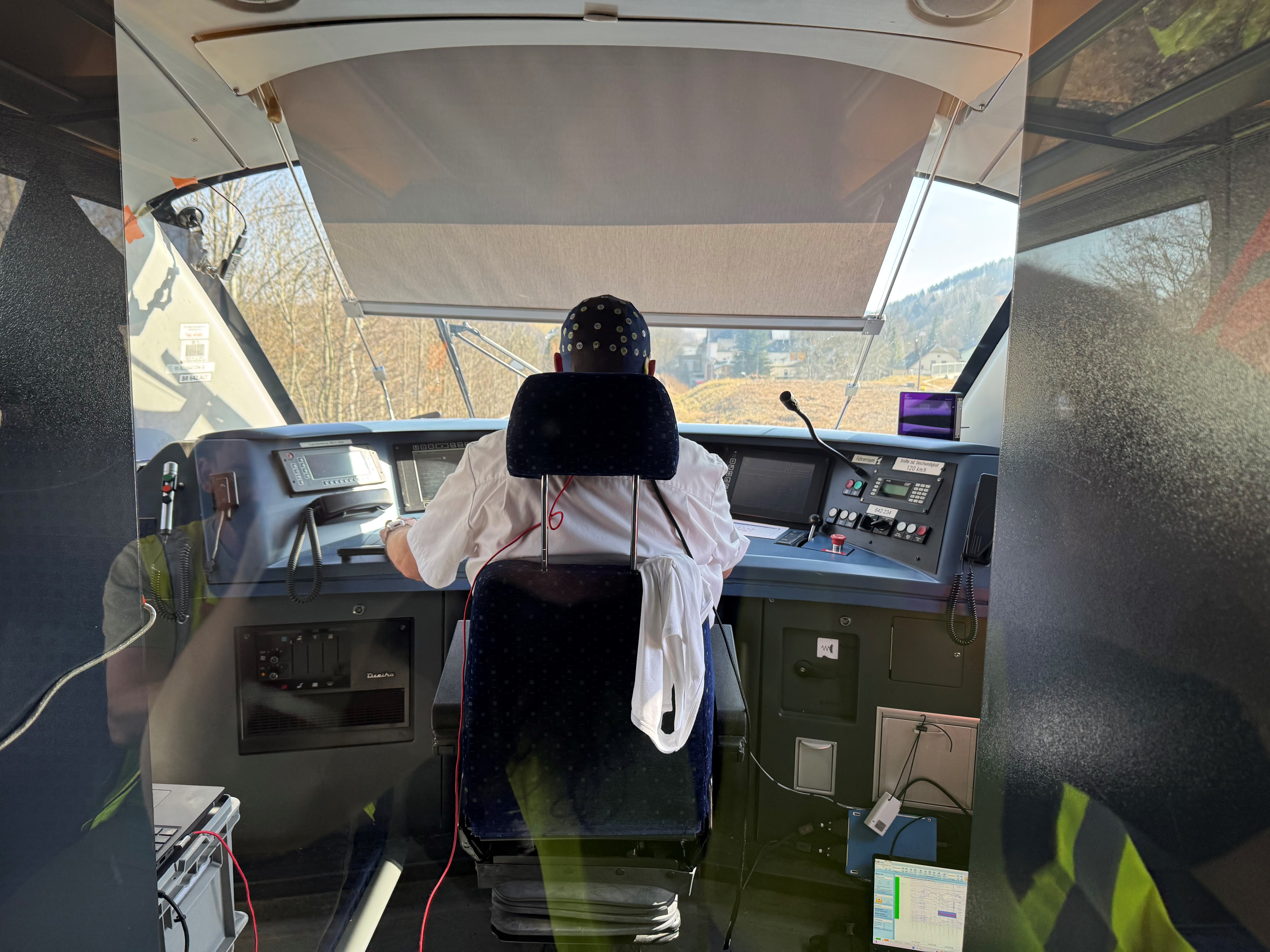}
    \caption{Real-world track.}
    \label{fig:envirs}
\end{figure}

The study took place in a simulator (July and August 2024, see Figure \ref{fig:first}) and a real-world environment (March 2025, see Figure \ref{fig:envirs}).

\subsection{Study design, Mental Fatigue Induction and Task Description}

The experimental session followed a standardized sequence in both the simulator and real-world environments, and took approximately two hours per participant.

Each session began with a 15-minute drive ('Drive 1') serving as a baseline phase without additional cognitive demands. This was followed by a one-hour MF induction task administered using PsychoPy (Version 2025.1.1). Participants completed an auditory n-back task in which three-digit numbers were presented aurally at regular intervals. Participants were required to respond by entering the target number on a numeric keypad. The task was adapted from \cite{hamann2026examining} and comprised four difficulty levels (0-back through 3-back), each presented twice in blocks of five minutes, with block order randomized across participants. Prior to the main task, participants completed a practice phase of the n-back task, continuing until they achieved an accuracy of at least 80\% on each difficulty level. Following the cognitive task, a second 15-minute drive ('Drive 2') was conducted to assess the effects of induced MF on physiological and performance measures.

To measure behavioral changes, a reaction time task was included. For this, in the simulator, eight orange, safety-vest-colored 2×~2\,m cubes were positioned at irregular intervals 2\,m from the track, while in the real-world environment, eight orange snow fences of comparable size and color were erected at the same 2\,m distance from the track. The color and contrast of the targets were matched as closely as possible across environments to approximate the brown-green background typical of the real-world track, ensuring comparable visual salience of the stimuli in both settings. Participants were instructed to sound the signal horn as soon as they detected the target, and the resulting reaction time was recorded as a behavioral measure of sustained attention.

While the scenario driven in the simulator consisted of the track from Braunschweig to Gifhorn in Lower Saxony, the real-world track was driven on the track from Schwarzenberg to Schlettau in Saxony. The two driving environments differed in their operational characteristics, reflecting the technical and regulatory constraints of each context. In the simulator, the train operated without the Punktförmige Zugbeeinflussung (PZB) system and with speed automatically regulated; participants were required only to continuously operate the dead man's handle and monitor the train's speed. There were 5 closed gated level crossings and 9 ungated level crossings. In the real-world environment, participants drove under Grade of Automation 1 (GOA1) conditions, performing standard operational tasks including speed regulation and managing level crossings: 4 closed crossings with gates, 10 ungated crossings, and one gated crossing that needed to be closed by manually by a gate attendant accompanying the driver on the train. While a fully automated driving condition analogous to the simulator could not be implemented in the real-world setting, the task design in both environments was intentionally aligned to approximate GOA2-level operation as closely as possible, minimizing differences in the active driving demands placed on participants.

\hfill

\subsection{Mental Fatigue Measures}

\subsection{Subjective Fatigue}
Subjective fatigue was assessed using the Karolinska Sleepiness Scale (KSS), a nine-point self-report scale on which participants rate their momentary level of sleepiness. Participants completed the KSS four times across the session: immediately before the first drive, and immediately after each of the first drive, n-back task, and second drive, allowing within-participant tracking of subjective sleepiness across the fatigue-induction protocol.

\subsubsection{Electroencephalography}
In the simulator, EEG was recorded using a Brain Products LiveAmp-32 system with active, gel-based Ag/AgCl electrodes and BrainVision Recorder (Version 1.26; Brain Products GmbH, Gilching, Germany) at a sampling rate of 500,Hz. Twenty-six electrodes were positioned according to the international 10--20 system with an online reference at FCz; six electrode positions (FT9, FT10, T7, T8, TP9, TP10) were omitted to accommodate the eye-tracking glasses. In the real-world driving condition, EEG was recorded using an ANT Neuro eego sports system with 64 sponge electrodes at a sampling rate of 500 Hz.

\subsubsection{Eye Tracking}
Blink duration was recorded using Pupil Core eye-tracking glasses (Pupil Labs, Berlin, Germany). 

\subsubsection{Cardiovascular, Respiratory, and Electrodermal Measures}
Heart rate variability was recorded using a Polar H10 chest strap (Polar Electro, Kempele, Finland). Breathing rate was recorded using a Vernier Go Direct Respiration Belt (Vernier Science Education, Beaverton, OR, USA). Electrodermal activity was recorded using a Shimmer3 GSR+ unit (Shimmer Sensing, Dublin, Ireland). 

\subsection{Behavioral Indicator}
Two behavioral indicators were used to assess vigilance and responsiveness during driving. First, the number of missed dead man's handle activations was recorded as an indicator of lapses in continuous manual responding. This was only possible in the simulator environment. Second, reaction time to visual targets placed alongside the track was measured in both environments.

After the study, participants rated each sensor's comfort level and their overall experience with the sensors on a scale from 1 (very uncomfortable) to 5 (very comfortable). 

\subsection{Procedure}
\subsubsection{Simulator study}

Upon arrival at the research institute, participants provided written informed consent by signing the data protection declaration and participant information sheet. The study procedure was then explained in detail, after which all physiological sensors were attached.
The simulator study was conducted in the RailSET (Railway Simulation Environment for Train Drivers and Operators; Figure \ref{fig:first}), a high-fidelity train driver's cab simulator located at the DLR Institute of Transportation Systems. The simulator features an original control panel from a real traction unit and uses VIRES software (VIRES Simulationstechnologie GmbH, Bad Aibling, Germany) as its simulation engine. The forward view is projected via a video projector, while lateral views are displayed on dedicated screens positioned at the side windows. Ambient cabin sounds modelled on a real train driver's cab are reproduced through an integrated audio system.

\subsubsection{Real-world study}
The real-world study took place on a railway track in the Schwarzenberg area in Saxony, Germany. Participants were equipped with all physiological sensors at Schwarzenberg station prior to the start of the session. The experimental procedure followed the same sequence as the simulator study: a 15-minute baseline drive, a one-hour n-back task completed while the train was stopped, and a second 15-minute drive. Following the final drive, participants returned to Schwarzenberg station, where the next participant was already waiting. To ensure safety, the track was closed to all other train traffic for the entire duration of the experimental sessions.

\subsection{Participants}

As part of their professional licensing requirements, all participating train drivers undergo regular medical fitness assessments in accordance with the German Triebfahrzeugführerscheinverordnung (TfV), which include mandatory testing of visual and auditory function. Participants can therefore be assumed to have had normal or corrected-to-normal vision and hearing at the time of testing.

A total of 14 professional train drivers participated in the simulator study (1 female, 13 male, $38 \pm 15$ years old), with a mean driving experience of $14 \pm 15$ years. Participants reported very high familiarity with PZB (M = 8.9 on a scale from 1 to 10), high familiarity with LZB (M = 7.2), and medium-to-low familiarity with ETCS (M = 3.9). The study was approved by the institute's ethics board (reference no. 6/24).

The real-world study was conducted with a separate sample of six professional train drivers (all male; age $52 \pm 9$ years), with a mean driving experience of $23 \pm 2$ years. Participants reported very high familiarity with PZB (M = 10), low familiarity with LZB (M = 2.6), and low familiarity with ETCS (M = 1.7). The study was approved by the institute's ethics board (reference no. 1/25).

\subsection{Data Processing}
\subsubsection{Physiological Data}
All EEG data were pre-processed in BrainVision Analyzer 2.2 (Brain Products GmbH, Gilching, Germany). In order to make both recordings comparable, only the 26 measurement locations used both in the simulated and real-world drive were pre-processed and analyzed. The data were down-sampled from 500 to 256 Hz and re-referenced to average reference. To remove noise, the data were filtered between 0.5 and 40 Hz using a 4th order IIR filter, with an additional notch filter at 50 Hz to remove remaining line noise. Ocular artifacts were corrected using an independent component analysis specifically trained for ocular data. Motion artifacts and other noise were identified and removed by means of semi-automatic inspection. 

A global average across a whole drive was computed to get a general impression of changes between Drive 1 and Drive 2. The data were further segmented into epochs of 2 s with 0.5 s overlap per block, then converted into the frequency domain using a Fast Fourier Transformation with a Hanning window with 10 \% overlap, and then averaged. The data were exported as Power Spectral Density (PSD) raw sum ($\mu V^2/\text{Hz}$) as described below.

We calculated three different frequency bands as possible measures of MF, which were frontal theta power (4-8 Hz) at the averaged electrodes Fz, F3, F4; parietal alpha power (8-13 Hz) at the averaged electrodes Pz, P3, P4; and parietal beta power (13-30 Hz) at the averaged electrodes Pz, P3, P4.

Electrodermal activity was processed per participant with NeuroKit2's eda\_process function (sampling rate: 121 Hz), decomposing the signal into tonic and phasic components. Skin conductance response peaks were identified using the function's default peak-detection algorithm and summed to obtain a peak count per participant.

R-peaks were detected from the ECG signal per participant using NeuroKit2's ecg\_process function (sampling rate: 130 Hz), from which inter-beat intervals (IBIs) were derived. IBIs outside 280–1500 ms or deviating more than 20\% from the preceding interval were flagged as artifacts and replaced via linear interpolation. HRV was then quantified as the root mean square of successive differences (RMSSD) computed over a 10-minute window centered on each beat, sliding across the recording.

Blink onset and offset events were extracted from the Pupil Core blinks.pldata files and paired chronologically into individual blink episodes. Blink duration was then computed as the difference between offset and onset timestamps for each episode.

Heart Rate Variability measure RMSSD, blinks, and breathing rate were z-score normalized within each participant.

\subsubsection{Behavioral Data}
Reaction times were calculated as the interval between the timepoint at which the visual target (cube or snow fence) first became visible to the participant and the timepoint at which the signal horn was pressed in response.

Missed dead man's handle data was extracted from the simulator logging file.

\subsection{Data Analysis}

In addition to descriptive results, we tested Bonferroni-corrected paired Wilcoxon-signed-rank tests per sensor and environment. Table \ref{tab:completeness} displays the available data per sensor and environment. Especially in the real-world drive, sensors failed due to USB connectivity failures caused by train vibration, Bluetooth saturation when several high-frequency sensors ran at once, and high data rates from the eye-tracking and camera systems.

\begin{table}[ht]
\centering
\caption{Available participant data per condition and sensor.}
\label{tab:completeness}
\vspace{6pt}
\begin{tabular}{lcc}
\toprule
Indicator & Simulator & Real-world \\
\midrule
EEG       & 14 & 6 \\
Blink duration       & 9 & 3 \\
Breathing rate       & 10 & 6 \\
SCR peaks  & 10 & 5 \\
RMSSD ($z$-score) & 10 & 5 \\
Reaction time & 13 & 3 \\
KSS & 14 & 6 \\
\bottomrule
\end{tabular}
\end{table}

\section{Results}

\begin{figure}
\centering
    \includegraphics[width=\textwidth]{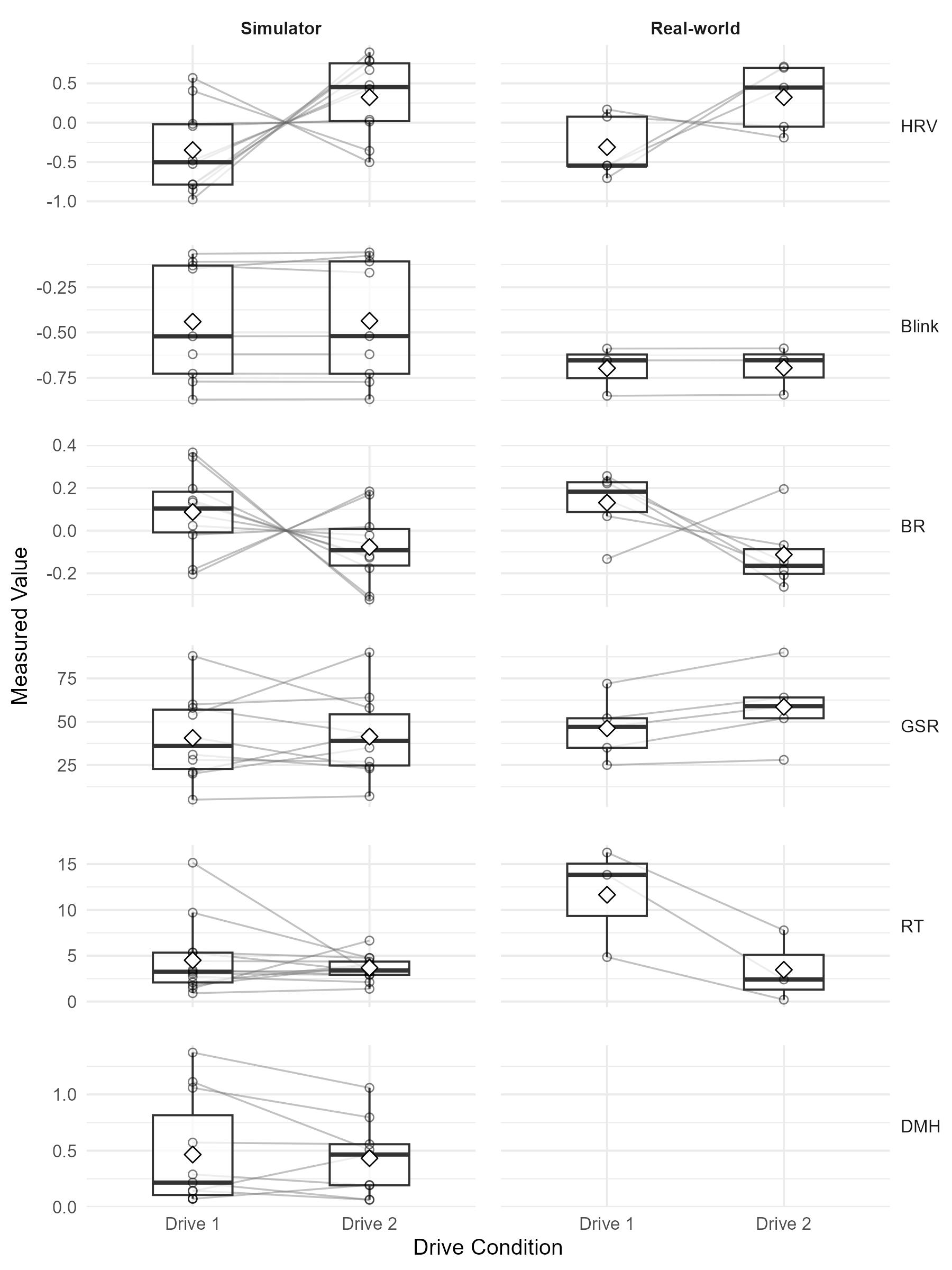}
    \caption{Sensor values per drive and environment. HRV = heart rate variability (RMSSD, z-scored); Blink = blink duration (ms); BR = breathing rate (z-scored); SCR = galvanic skin response peaks (n); RT = reaction time (ms); DMH = dead man's handle missed activations (n).}
    \label{fig:othersensors}
\end{figure}

\begin{figure}
\centering
    \includegraphics[width=\textwidth]{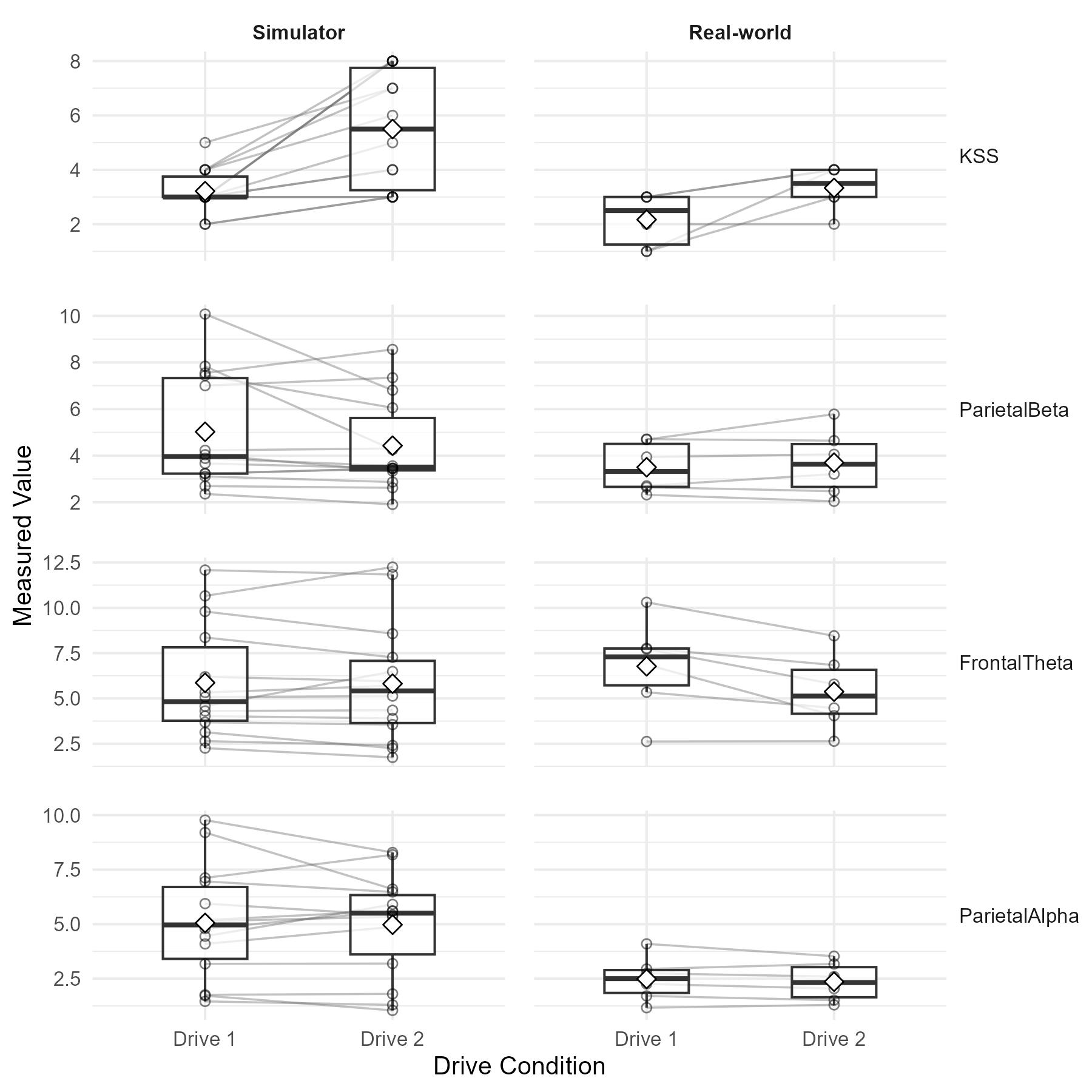}
    \caption{Sensor values per drive and environment. KSS = Karolinska Sleepiness Scale.}
    \label{fig:eegresults}
\end{figure}

None of the paired Wilcoxon-signed-rank tests were significant. In the following, we report descriptive data for each sensor. It is also displayed in Figures \ref{fig:othersensors} and \ref{fig:eegresults}.

\subsection{Subjective Fatigue}
Karolinska Sleepiness Scale (KSS) ratings increased from the first to the second drive in both environments. In the simulator, mean sleepiness increased from 3.21 (SD = 0.80) before the n-back task to 5.50 (SD = 2.14) after, indicating a notable rise in subjective sleepiness across the session. In the real-world environment, KSS ratings were lower overall, with a mean of 2.17 (SD = 0.98) before and 3.33 (SD = 0.82) after the task. In both environments, the increase in sleepiness from Drive 1 to Drive 2 suggests that the n-back task successfully induced MF. Notably, simulator participants reported higher baseline sleepiness and a larger absolute increase compared to the real-world sample, though the smaller sample size in the real-world condition (n = 6 vs. n = 14) limits direct comparison.

\subsection{EEG}
Frontal theta power decreased from Drive 1 to Drive 2 in the real-world environment, from a mean of 6.77 (SD = 2.59) to 5.37 (SD = 2.09). In the simulator, frontal theta remained largely stable across drives, with means of 5.87 (SD = 3.13) in Drive 1 and 5.81 (SD = 3.28) in Drive 2. Parietal alpha power showed a similar pattern of stability or slight decrease in both environments, with means of 5.06 (SD = 2.60) in Drive 1 and 4.97 (SD = 2.33) in Drive 2 in the simulator, and 2.49 (SD = 1.03) in Drive 1 and 2.36 (SD = 0.90) in Drive 2 in the real environment. Parietal beta power decreased from Drive 1 (M = 5.02, SD = 2.43) to Drive 2 (M = 4.43, SD = 1.98) in the simulator, while showing a slight increase from Drive 1 (M = 3.50, SD = 1.08) to Drive 2 (M = 3.70, SD = 1.40) in the real environment. Across all three frequency bands, none of the observed changes were in the hypothesized direction of increasing power with fatigue.

\subsection{Heart Rate Variability (RMSSD)}

HRV showed a consistent pattern across both environments: values were lower during Drive 1 and higher during Drive 2. In the simulator, mean z-scored RMSSD was -0.35 ($SD = 0.55$) in Drive 1 and 0.32 (SD = 0.50) in Drive 2. A similar direction was observed in the real environment, with means of -0.31 (SD = 0.40) and 0.32 (SD = 0.42) respectively. This pattern suggests relatively higher parasympathetic activity, and thus lower physiological arousal, during Drive 2 compared to Drive 1, in both settings.

\subsection{Breathing Rate}
Breathing rate (z-scored) followed a comparable pattern in both environments, with slightly elevated values during Drive 1 and reduced values during Drive 2. In the simulator, mean breathing rate was 0.09 (SD = 0.19) in Drive 1 and -0.08 (SD = 0.17) in Drive 2. In the real environment, values were 0.13 (SD = 0.15) and -0.11 (SD = 0.16) respectively. 

\subsection{Electrodermal Activity}
In the simulator, SCR counts were nearly identical across conditions (Drive 1: M = 40.6, SD = 24.57; Drive 2: M = 41.4, SD = 24.10), indicating no meaningful difference. SCR peaks showed somewhat higher counts during Drive 2 compared to Drive 1 in the real environment (Drive 1: M = 46.2, SD = 17.85; Drive 2: M = 58.6, SD = 22.33), suggesting greater sympathetic activation during Drive 2 in that setting.  Variability was considerable across participants in both environments.

\subsection{Blink Duration}
Blink duration (z-scored) was negative across all conditions, indicating that participants' blink durations were generally below their individual baseline. Values were very similar between drives 1 and 2 within each environment; in the simulator, Drive 1: M = -0.44 (SD = 0.33) and Drive 2: M = -0.44 (SD = 0.33); in the real environment, Drive 1: M = -0.70 (SD = 0.14) and Drive 2: M = -0.70 (SD = 0.13). The near-identical values between drive conditions suggest blink duration was not differentially sensitive to the drive manipulation in this dataset. Notably, blink durations were more negative in the real environment than in the simulator, which may reflect higher overall alertness or visual engagement in real-world driving.

\subsection{Reaction Time}
Reaction times were higher in Drive 1 than Drive 2 across both environments. In the simulator, mean reaction time was 4.51 s (SD = 3.93) in Drive 1 and 3.68 s (SD = 1.33) in Drive 2. In the real environment, this difference was more pronounced: Drive 1 M = 11.65 s (SD = 6.01) versus Drive 2 M = 3.46 s (SD = 3.90), though the real-environment sample was very small (n = 3) and results should be interpreted with caution.

\subsection{Dead Man Handle Reminders}
This measure was only available for the simulator condition. Missed reminders were slightly more frequent in Drive 1 (M = 0.47, SD = 0.49) than Drive 2 (M = 0.43, SD = 0.34), though the distributions overlapped considerably (Drive 1 median = 0.22, Drive 2 median = 0.47), and variability was high across participants.

\subsection{Sensor comfort}
\begin{figure}
\centering
\includegraphics[width=\textwidth]{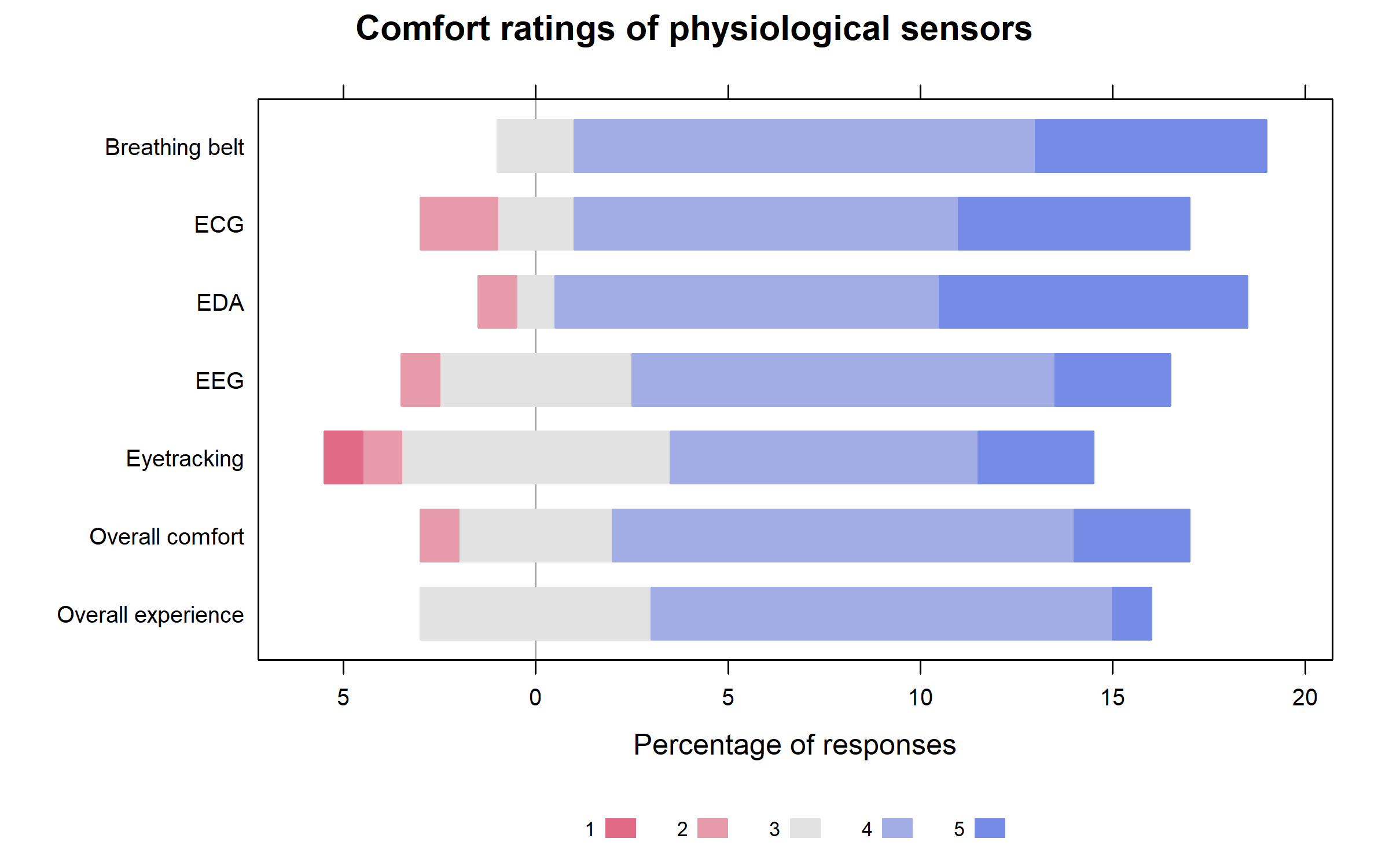}
\caption{Each sensor was rated on a Likert scale from 1 = not at all comfortable to 5 = very comfortable. Data for simulator and real environment study together.}
\label{fig:comfort}
\end{figure}

Across all sensors, median ratings were consistently 4 (IQR = 1.00 for most sensors), with EDA (M = 4.25, SD = 0.79) and the breathing belt (M = 4.20, SD = 0.62) rated as most comfortable, and eye tracking rated as the least comfortable (M = 3.55, SD = 1.00), while overall sensor comfort (M = 3.85, SD = 0.75) and overall experience (M = 3.74, SD = 0.56) were both rated favorably on average, see Figure \ref{fig:comfort} 

\clearpage

\section{Discussion}

The present study investigated multiple subjective, physiological, and behavioral indicators of MF in professional train drivers across a high-fidelity simulator and a real-world rail environment. We return here to the four research questions guiding the study, and to the hypotheses formulated for each indicator.

\subsection{RQ1: Fatigue Induction}
Our first question asked whether subjective fatigue was rated consistent with increasing MF. This was the case. KSS ratings increased from Drive 1 to Drive 2 in both environments, supporting the hypothesis that self-reported fatigue would rise following the induction task. In the simulator, the increase was more pronounced (from M=3.21 to M=5.50). In the real-world environment, the change was more modest (from M=2.17 to M=3.33). This difference should be interpreted with caution. Simulator participants reported notably higher sleepiness even before the session began. This may reflect the enclosed, passive nature of the simulator cab, or accumulated travel fatigue prior to arrival. A lower starting point in the real-world group, maybe due to the engaging nature of the environment, could partly explain the smaller absolute increase, rather than a weaker effect of the fatigue induction task itself.

\subsection{RQ2: Indicator Sensitivity}
Our second question asked which physiological measures would show sensitivity to fatigue-related changes. Here the results were mixed, and diverged from our hypotheses in most cases.

\subsubsection{Heart Rate Variability and Breathing Rate}
HRV and breathing rate were the two indicators that matched their hypothesized effects most clearly. We had expected MF to be associated with higher HRV \cite{matuz2021enhanced} and slower breathing \cite{grassmann2016respiratory}. Both patterns emerged, and did so consistently across both environments. RMSSD was lower during Drive 1 and higher during Drive 2. Breathing rate showed the inverse pattern. These directions are consistent with theoretical expectations. Increased parasympathetic tone and slower respiration are associated with reduced arousal and passive fatigue states \cite{lal2001critical, charles2019measuring}. The convergence of two independent autonomic indicators, across two environments with very different sample sizes and operational demands, strengthens confidence in this pattern. This is consistent with the broader autonomic monitoring literature, which has identified HRV as a comparatively robust indicator of operator state, resilient to motion artifacts and well suited to wearable integration \cite{BrookhuisDeWaard2010}. Our findings extend this evidence specifically to train driving. From a practical standpoint, HRV and breathing rate can be captured unobtrusively with chest-worn sensors and showed directionally consistent results even in a small real-world sample. We recommend that future driver monitoring research and system development prioritize these two indicators.

\subsubsection{Electrodermal Activity}
We had hypothesized that MF would be associated with fewer SCR peaks \cite{posada2017sleep}. This was not supported. SCR peaks increased slightly in the real-world environment from Drive 1 to Drive 2, and were essentially unchanged in the simulator. The direction observed in the real-world sample is opposite to what was expected. It may reflect greater situational demands during Drive 2, for instance route familiarity reducing attentional engagement, or subtle differences in the operational context of the second drive. Given the very small real-world sample (n=5), it may also reflect individual variability rather than a systematic effect. The absence of change in the simulator suggests that EDA was not sensitive to the fatigue manipulation under the conditions tested here. High inter-individual variability in both environments is consistent with the general EDA literature, which consistently reports large individual differences in tonic and phasic response patterns \cite{BrookhuisDeWaard2010}.

\subsubsection{EEG}
We had hypothesized increased frontal theta and parietal alpha power with fatigue based on Hamann et al. \cite{hamann2023assessing}. Neither hypothesis was supported. Frontal theta on F3 and F4 decreased rather than increased between drives. One possible explanation is that the repeated task switching across the three experimental phases (Drive 1, n-back task, Drive 2) increased alertness beyond what the study design anticipated. More broadly, this result fits a growing body of evidence that EEG-based fatigue markers are highly context-sensitive and transfer poorly across tasks, populations, and measurement conditions \cite{Dehais2019}. We treat this non-replication as a meaningful finding in its own right. It highlights the gap between classification performance under idealized laboratory conditions and the interpretability of EEG under realistic operational constraints.

\subsubsection{Blink Duration}
We had hypothesized longer blink duration with fatigue \cite{bafna2021mental}. This was not supported. Blink duration was negative across all conditions, meaning durations stayed below individual baseline throughout. Values were nearly identical between Drive 1 and Drive 2 within each environment. This null result sits alongside the EEG non-replication in an interesting way. Prior work has shown that EEG and oculomotor indicators of vigilance often move together \cite{bodala2016eeg}. In our data, both modalities were similarly insensitive to the fatigue manipulation, which is at least internally consistent even if it does not support either hypothesis. One possible explanation is that participants remained visually engaged throughout both drives, suppressing blink duration regardless of fatigue state. This would match findings that task-relevant visual demands can override fatigue-related blink lengthening \cite{hu2021exploration}. It is also worth noting that we measured blink duration in isolation, rather than a composite index such as PERCLOS. Reviews of ocular drowsiness measures suggest no single index is sufficient on its own \cite{abe2023perclos}, so a composite approach may have been more sensitive here. Blink durations were also more negative in the real-world environment than in the simulator, which may indicate higher overall visual alertness during real-world driving, consistent with the lower KSS ratings observed in that group.

\subsubsection{Reaction Time and Dead Man's Handle}
We had hypothesized longer reaction times and more missed dead man's handle activations with fatigue. Neither was supported. Reaction times were faster, not slower, in Drive 2 across both environments. The most likely explanation is a practice effect. Participants became more familiar with detecting the orange targets and responding with the signal horn over the session, which may have offset or masked any fatigue-related slowing. This interpretation is supported by the especially large improvement in the real-world environment, where the task was entirely novel. Future studies should include a longer familiarization phase for the reaction time task. Another explanation could be over-compensation when becoming fatigued \cite{dorrian2007simulated}, where operators are aware of strong fatigue and actively increase their efforts to perform well at their task. Dead man's handle miss rate in the simulator was very low and showed no meaningful change between drives. This suggests the measure lacks sensitivity once the response has become highly automated, a well-documented limitation of vigilance monitoring devices in general \cite{BrookhuisDeWaard2010}.

\subsection{RQ3: Ecological Robustness}
Our third question asked which indicators remain stable and interpretable under real-world rail conditions. HRV and breathing rate again stand out. Both showed the same directional pattern in the small, technically constrained real-world sample as in the controlled simulator. This is a meaningful form of robustness, since real-world data collection introduced substantial technical challenges. These are important practical lessons for future field studies. They point to a need for purpose-built data acquisition pipelines for operational rail environments, and caution against transferring laboratory sensor setups to the field without systematic robustness testing beforehand.

The simulator and real-world conditions also differed in sample size, participant composition, timing of data collection, and operational task demands. This limits direct comparison between environments. The present study is best understood as a feasibility investigation. It shows that the sensor system was deployable in both contexts, and that consistent patterns emerged for at least some indicators despite these constraints. The small sample sizes, particularly in the real-world study (n=6), preclude inferential statistics for most measures. All reported patterns beyond the EEG analyses should be treated as descriptive and exploratory. Replication with larger samples and a more controlled between-environment design is needed before firm conclusions can be drawn about environment-specific differences.

\subsection{RQ4: Practical Feasibility}
Our fourth question asked how drivers perceived the comfort and usability of the sensors used. Ratings were favorable overall, with a median of 4 out of 5 for most sensors. EDA (M=4.25, SD=0.79) and the breathing belt (M=4.20, SD=0.62) were rated as most comfortable. Eye tracking was rated as least comfortable (M=3.55, SD=1.00). This result is worth reading together with the sensitivity findings above. The sensor rated most comfortable, the breathing belt, was the one that also showed the clearest and most robust fatigue-related patterns. Eye tracking, meanwhile, was both the least comfortable sensor and the one whose blink-duration output showed no fatigue sensitivity in our data. This convergence strengthens the case for prioritizing chest and abdomen worn autonomic sensors in future operational monitoring systems, both on diagnostic and on practical grounds.

\subsection{Returning to the Ironies of Automation}
The Introduction framed this study in terms of Bainbridge's Ironies of Automation, the observation that increasing automation reduces active engagement while still requiring reliable monitoring \cite{BAINBRIDGE1983129}. Our behavioral results illustrate this irony directly. Reaction times improved and dead man's handle misses stayed flat across the session, even though subjective sleepiness rose. Drivers can apparently maintain overt task performance through practiced, automatized responses, even as their underlying state shifts. This is precisely the failure mode that motivated our multisensor approach. Autonomic measures, unlike behavioral output, do not depend on a driver actively producing a correct response. They may therefore offer a more direct route to detecting the kind of vigilance erosion that behavioral monitoring, including traditional dead man's handle systems, is known to miss \cite{whitlock2017driver, foot2008questions, berdal2024towards}.

\subsection{Adequacy of the Fatigue Induction}
A cross-cutting issue affecting most null results is whether the experimental procedure induced a level and type of fatigue sufficient to produce measurable physiological change within a 15-minute post-task drive. The KSS data suggest subjective fatigue did increase, but the magnitude, particularly in the real-world environment, was modest. The n-back task is a standard laboratory induction tool, but it produces active cognitive fatigue through sustained effortful processing. The fatigue most relevant to GoA2 monitoring arises instead from prolonged inactivity and underload. These two forms of fatigue may differ not only in experience but also in their physiological signatures \cite{desmond2000active, brandenburger2021task}. A more ecologically valid induction, closer in duration and character to a real monitoring shift, would target mental fatigue more directly and might produce stronger physiological effects.

\subsection{Practical Implications}
Taken together, these results suggest that HRV and breathing rate are the most promising starting point for fatigue monitoring in rail operations, based on current evidence. Both showed consistent directional changes across environments, are technically robust in wearable form, and were rated among the most comfortable sensors by drivers themselves. EEG, despite its theoretical appeal and strong laboratory evidence base, showed no consistent fatigue-related pattern here and faces substantial practical challenges in field deployment. EEG, EDA, blink duration, and the behavioral measures tested here were not sensitive to the manipulation under the conditions studied, though methodological factors, such as induction type and measurement duration, may account for part of this insensitivity rather than a genuine absence of effect.

Future work should prioritize ecologically valid fatigue induction paradigms, larger samples, and purpose-built data acquisition infrastructure for operational rail environments. Sensor comfort and wearability, addressed here through Likert ratings, merit further systematic investigation, since long-term acceptance by train drivers is a prerequisite for any operational monitoring system.

\section{Conclusion}
The present study examined the feasibility and sensitivity of a multisensor approach to MF assessment in professional train drivers across a simulator and a real-world rail environment. Subjective sleepiness increased across the session in both settings, confirming that the fatigue induction procedure was effective at least at the self-report level. Of the physiological indicators examined, heart rate variability and breathing rate showed the most consistent and theoretically coherent patterns, remaining stable across both the controlled simulator and the technically constrained real-world environment. Both sensors were also rated among the most comfortable by drivers, supporting their use as a pragmatic foundation for driver state monitoring in rail and their integration into wearable sensing systems.

The remaining indicators, including EEG, electrodermal activity, blink duration, and behavioral measures, did not show clear fatigue-related changes. The failure to replicate expected frontal theta increases underscores the context-sensitivity of neurophysiological fatigue markers and cautions against their uncritical transfer across operational domains. Reaction times and dead man's handle performance improved or stayed flat despite rising subjective fatigue, illustrating in practice the Ironies of Automation that motivated this study: drivers can sustain overt task performance through practiced, automatized responses even as their underlying state shifts. This gap between behavioral output and internal state is precisely what autonomic monitoring is well placed to address. Real-world data collection additionally revealed practical challenges related to vibration, sensor connectivity, and concurrent high-frequency data acquisition that must be addressed in future field deployments.

Future work should prioritize ecologically valid fatigue induction paradigms that better approximate the passive monitoring demands of GoA2 operation, larger samples, and purpose-built acquisition infrastructure for operational rail environments. This study provides applied, ecologically grounded evidence on which indicators hold promise, how drivers experience the associated sensors, and what methodological conditions are needed to advance fatigue monitoring in increasingly automated rail systems.

\section{Acknowledgements}
Funded under grant number 03WIR1214B by the German Federal Ministry of Education and Research (BMBF) within the "WIR! – Wandel durch Innovation in der Region" funding initiative. We thank Prof. Lewis Chuang, Dr. Giuseppe Sanseverino and Ferenc Rozsa for help with the data acquisition in the real-world drives. 

\section{Declaration of generative AI and AI-assisted technologies in the writing process.}

Statement: During the preparation of this work, the authors used Claude Sonnet 5 (Version July 2026) to support spell-checking and refine wording. After using this tool, the authors reviewed and edited the content as needed and take full responsibility for the content of the published article.

\clearpage


 \bibliographystyle{elsarticle-num} 
 \bibliography{02_references}





\end{document}